\documentclass{ws-mpla}
\usepackage{cite}
\usepackage{graphicx}
\usepackage{xcolor}
\usepackage{float}
\usepackage{amsmath,amsfonts,amssymb}
\usepackage{subfig,ragged2e}
\usepackage{slashed}
\usepackage{hyperref}

\begin{document}


\catchline{}{}{}{}{}

\title{Probing and Knocking with Muons}
\author{Leyun Gao (1), Cheng-en Liu (1), Qite Li (1), Chen Zhou (1), Qiang Li (1), Liangwen Chen (2), Xueheng Zhang (2), Yu Xu (2), and Zhiyu Sun (2)}
\address{(1) State Key Laboratory of Nuclear Physics and Technology, School of Physics, Peking University, Beijing, 100871, China; \\ (2) Institute of Modern Physics, Chinese Academy of Science, Lanzhou, 730000, China;\\University of Chinese Academy of Sciences, Beijing, 100049, China}
\maketitle

\begin{abstract}
We propose here a set of new methods involving probing and knocking with muons (PKMu). There is a wealth of rich physics to explore with GeV muon beams. Examples include but not limited to: muon scattering can occur at large angles, providing evidence of potential muon-philic dark matter or dark mediator candidates; muon-electron scattering can be used to detect new types of bosons associated with charged lepton flavor violation; precise measurements of GeV-scale muon-electron scattering can be employed to probe quantum correlations.
\keywords{muon physics; beyond the Standard Model; quantum entanglement.}
\end{abstract}
\ccode{PACS Nos.: 14.60.Ef, 12.60.-i, 03.65.Ud}

\section{Introduction}

Among all elementary particles, the muon has a unique role in bridging fundamental research with applied studies, albeit to a lesser extent than its cousin, the electron. Both cosmic and man-made muons serve as powerful tools for exploring the mysteries of our universe and the unknown. For instance, while the Standard Model (SM) of particle physics has proven remarkably successful, it falls short in explaining certain experimental observations, such as neutrino masses and dark matter (DM). A range of theories beyond the Standard Model has been proposed to tackle these challenges, introducing hypothetical particles and novel interactions. Understanding the nature of dark matter is currently a focal point for both cosmology and particle physics. The paradigm of Weakly Interacting Massive Particles (WIMPs) is extensively studied in the dark sector, with numerous experiments conducted to search for them. However, the lack of observed WIMPs has prompted the exploration of other theoretically motivated scenarios, including low-mass or muon-philic dark matter (MPDM)~\cite{Essig:2022dfa,Harris:2022vnx,Bai:2014osa}.

Muon scattering experiments are relatively scarce on an international scale. On one hand, experiments conducted in Europe and the United States from the 1960s to the 1980s~\cite{Drees:1983pd} primarily focused on detecting nuclear properties. At that time, the Standard Model had not yet been fully established, so the use of muons to search for new physics, such as charged lepton flavor violation (CLFV) or dark matter, was not yet fully developed. On the other hand, in recent decades, the investigation of new physics through muons has gained momentum. Notable examples include the anomalous magnetic moment experiment in the United States and experiments like MEG~\cite{MEG:2016leq} and Mu3e~\cite{Hesketh:2022wgw} in Europe, which focus on anomalous muon decay. These studies typically utilize free muons or muon decays. In recent years, however, experiments such as CERN's NA64~\cite{NA64:2024klw} and the planned MUonE~\cite{CarloniCalame:2015obs} have begun to reintroduce high-energy muons (approximately 160 GeV) for target scattering, exploring new physics and key Standard Model observables through muon scattering off nucleons or electrons. Despite this progress, muon scattering presents a wealth of untapped physical questions, and varying muon beam energies can be sensitive to different domains of new physics.

\begin{figure}
\centering
\includegraphics[width=1.\columnwidth]{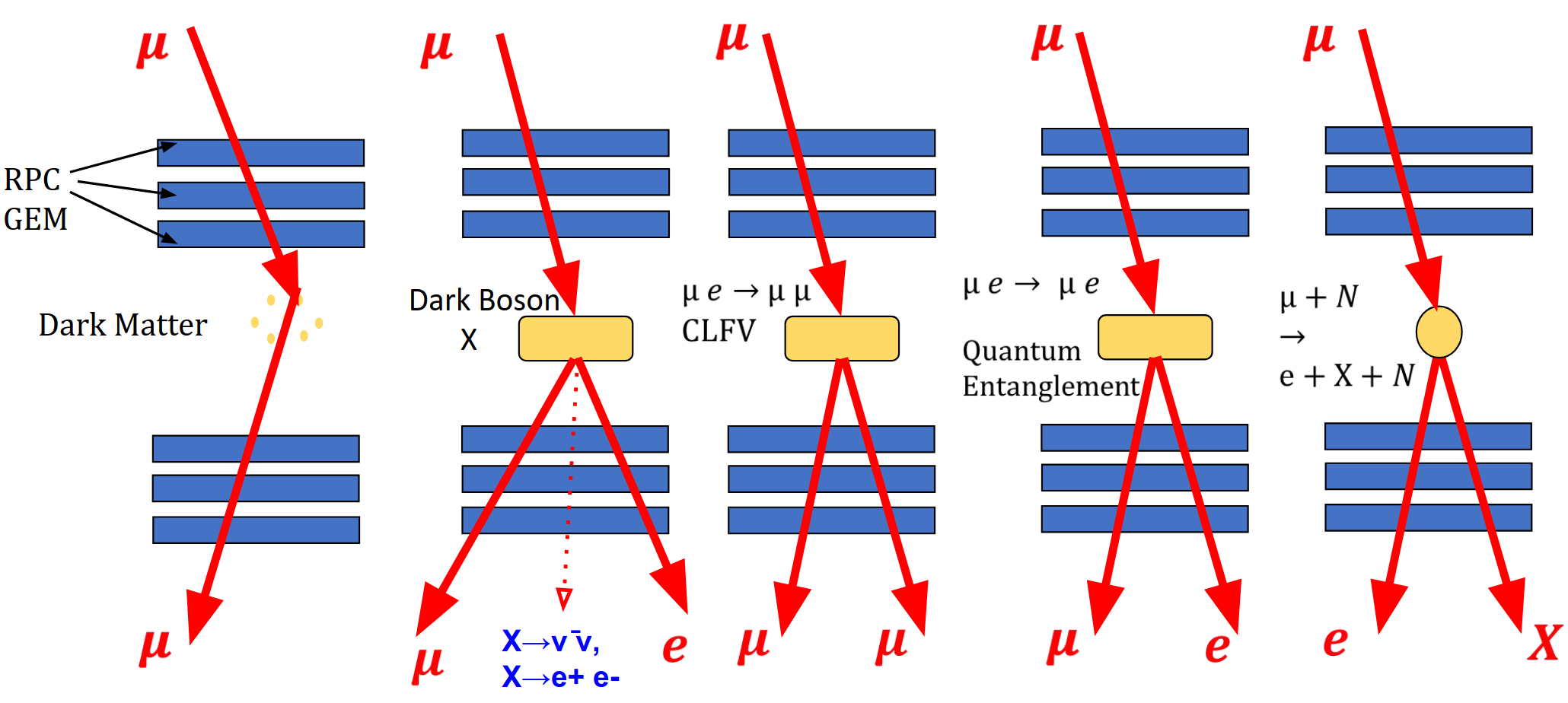}
\caption{\justifying Illustration of the PKMu proposal through probing and knocking with muons to investigate dark matter, dark bosons, charged lepton flavor violation, and quantum entanglement.}
\label{fig:pkmu}
\end{figure}

Historically, China has not conducted muon physics experiments due to a lack of suitable high-intensity proton and heavy ion accelerator facilities. However, this situation has changed significantly in recent years. For instance, the High Energy Fragment Separator beamline at the HIAF facility is expected to produce and deliver a high-intensity GeV-energy muon flux in the coming years~\cite{Xu:2025spd}. This opens up exciting opportunities for exploring rich physics with GeV muon beams. As illustrated in Fig.~\ref{fig:pkmu} and elaborated below, muon scattering can occur at large angles, revealing the potential of new types of dark matter~\cite{Ruzi:2023mxp,Yu:2024spj}; muon-electron scattering can be employed to detect new bosons associated with charged lepton flavor violation~\cite{Gao:2024xvf}; precise measurements of GeV-scale muon-electron scattering can also be utilized for probing quantum correlations~\cite{Gao:2024leu,Gao:2025kdi}.

Drawing on local expertise and the cost-effectiveness of performance, we plan to utilize Gas Electron Multiplier (GEM)~\cite{gemd} and Resistive Plate Chamber (RPC)~\cite{Riegler:2004gc} technologies for muon tracking. GEM is a well-established amplification technique that enables precise position detection of charged particles in gaseous detectors. It has been extensively employed in various high-energy physics experiments, including the CMS experiment~\cite{Abbas:2022fze,Pellecchia:2022lsd}. Similarly, RPCs have been widely utilized in the field of high-energy physics as particle detectors for nearly 50 years. Their popularity stems from several advantages, including a simple and robust structure, long-term stability, excellent timing resolution, ease of maintenance, and low cost. These characteristics make both GEM and RPC technologies highly suitable for our muon tracking efforts.

\section{Model-irrelative direct detection of muon-philic dark matter}

\begin{figure}
\centering
\includegraphics[width=1.\columnwidth]{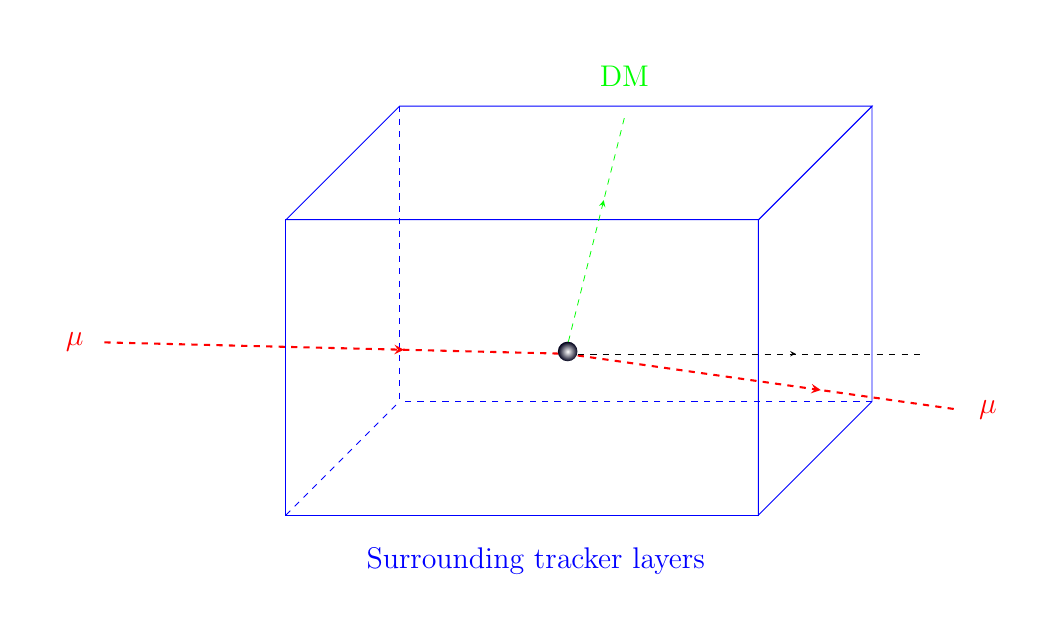}
\caption{\justifying Illustration of the experiment designed to detect muon-philic dark matter through interactions with free leptons. The spatial shifts of the leptons, resulting from their scattering with the dark matter, can be accurately measured using tracking detectors positioned around a vacuum region.}
\label{fig:muonbox}
\end{figure}

The focus of this section is to leverage muon detectors, specifically Resistive Plate Chambers (RPCs) and Gas Electron Multipliers (GEMs), for model-irrelative muon-philic dark matter searches. As an initial step, illustrated in Fig.~\ref{fig:muonbox}, we propose to utilize free cosmic muons interacting with dark matter within a designated volume surrounded by tracking detectors. This enables tracing potential interactions between muons and dark matter, facilitating the exploration of the elusive properties of the dark matter.

Simulation of the detector response is performed by \textsc{Geant4}~\cite{GEANT4:2002zbu}. A schematic design of the triple-GEM detector is available in Ref.~\cite{gemd}, and additional details can be found in Refs.~\cite{Yu:2024spj,Ruzi:2023mxp}. In the presence of surrounding dark matter, atmospheric or accelerator muons can experience spatial shifts, which serves as a distinctive signal feature. According to the Standard Halo Model~\cite{Lewin:1995rx,Aalbers:2018mfc}, the velocity distribution of dark matter particles in the galactic rest frame can be characterized by a Maxwell-Boltzmann distribution. For simplicity, we have also considered a scenario where the velocity distribution of dark matter is approximated as a constant value of $v = 220\ \mathrm{km/s}$. The differences between these two approaches are found to be minor.

In our study, we investigate a model-independent elastic scattering process between muons and dark matter, adhering to the principles of Newtonian mechanics. We obtain the kinematic spectrum of incident cosmic muons at sea level using the CRY program~\cite{CRY}, which yields a mean muon energy of 3–4 GeV. In this scenario, dark matter is treated as quasi-frozen in comparison to the rapidly incoming muons. Noticeably, this setup is complementary to many other dark matter search experiments, such as XENON1T and PandaX, where nuclei serve as quasi-static targets and dark matter as incoming particles.

In our cosmic muon-dark matter scattering experimental proposal, the event rate can be estimated as:
\begin{equation}
\mathrm dN/\mathrm dt=\rho V/M_\mathrm{DM} \times \sigma_{\mu,\mathrm{DM}} \times F_{\mu},
\end{equation}
where $F_{\mu}$ denotes the muon flux, which is approximately $1/60$~s$^{-1}$cm$^{-2}$ at sea level; the local density of dark matter is on the order of $\rho \sim 0.3$~GeV/$\rm{cm}^3$; and the division by $M_\mathrm{DM}$ gives the number of dark matter particles in volume $V$. For a dark matter mass of $M_{\mathrm{DM}} \sim 1$ (0.1, 10)~GeV and a detector volume of $V \sim 1\ {\rm m}^3$, the number of dark matter particles is estimated to be $3 \times 10^{5(6,4)}$. And the resulting sensitivity of the muon-DM scattering cross section for a one-year run is expected to be:
\begin{equation}\label{eq:muonbox}
\sigma_{\mu,\mathrm{DM}} \sim 10^{-12(-13,-11)}~{\rm cm}^2.
\end{equation}

Our methods offer advantages over certain specific \emph{exotic} dark matter scenarios~\cite{McKeen:2023ztq}, in which strongly interacting dark matter can accumulate in significant quantities within the Earth and be significantly slowed down through scattering with matter in the atmosphere or the Earth before reaching the detector target. In such scenarios, the dark matter number density can be $\sim 10^{15}$ cm$^{-3}$, and the sensitivity to dark matter and muon scattering cross sections can approach the microbarn level. However, it is important to note that detector acceptance and efficiency must be taken into account, as these factors can alter the estimated results. Further details are provided in Ref.~\cite{Yu:2024spj}.

\section{Probing light dark bosons}

Minimal scenarios featuring light (sub-GeV) dark matter that achieves its relic density through thermal freeze-out necessitate the introduction of an additional dark sector (DS), which includes light mediators. A particularly compelling example is the introduction of a new `dark' massive vector gauge boson mediator~\cite{Battaglieri:2017aum}. Many of these scenarios also propose the existence of new feeble interactions with muons, mediated by scalar, pseudoscalar, or vector-like particles.

A simple yet effective class of models is grounded in the $L_\mu - L_\tau$ gauge theory~\cite{Foot:1990mn, He:1990pn, He:1991qd, Foot:1994vd}, where $L_\mu$ and $L_\tau$ represent the lepton numbers of the muon and tau particles, respectively. The new gauge symmetry corresponds to a massive dark gauge boson, referred to as the $L_\mu - L_\tau$ gauge boson or $Z^\prime$.

We have preliminarily investigated the search potential for the $Z^\prime$ boson through muon on-target experiments, utilizing the process $\mu e^- \to \mu e^- X$, where $X$ decays invisibly. The expected experimental signature includes scattered muons and electrons from the target, emerging at large angles relative to the background noise, and simultaneously, minimal activity is anticipated in the sub-detectors located downstream of the interaction point. This proposal can be implemented on the 1--10 GeV muon beam from the HIAF-HFRS facility~\cite{Xu:2025spd}, which will be ready in 2025--2026. In comparison to existing experiments or proposals utilizing the 160 GeV muon beam at the CERN MUonE, we find that our approach offers significantly enhanced sensitivity in the $Z^\prime$ mass range around 10 MeV~\cite{wang2025}.

\section{Probing Charged Lepton Flavor Violation}

In the Standard Model, charged lepton flavor violation (CLFV) processes are extremely rare and challenging to observe experimentally. However, in many new physics scenarios, CLFV can be significantly enhanced, providing a sensitive probe for testing the Standard Model and exploring new physics. One such model introduces an additional $U(1)$ gauge symmetry, which corresponds to a massive neutral gauge boson denoted as $Z'$. For instance, Refs.~\cite{Ding:2024zaj,Langacker:2000ju} suggest that the additional $Z'$ current possesses the same gauge coupling and chiral structure as the Standard Model $Z$ boson, while also permitting charged lepton flavor violations, which can be quantified by
\begin{equation}
\lambda =\left(\begin{array}{lll}
\lambda_{e e} & \lambda_{e \mu} & \lambda_{e \tau} \\
\lambda_{\mu e} & \lambda_{\mu \mu} & \lambda_{\mu \tau} \\
\lambda_{\tau e} & \lambda_{\tau \mu} & \lambda_{\tau \tau}
\end{array}\right).
\label{eq:lambda}
\end{equation}

\begin{figure}
\centering
\includegraphics[width=1.\columnwidth]{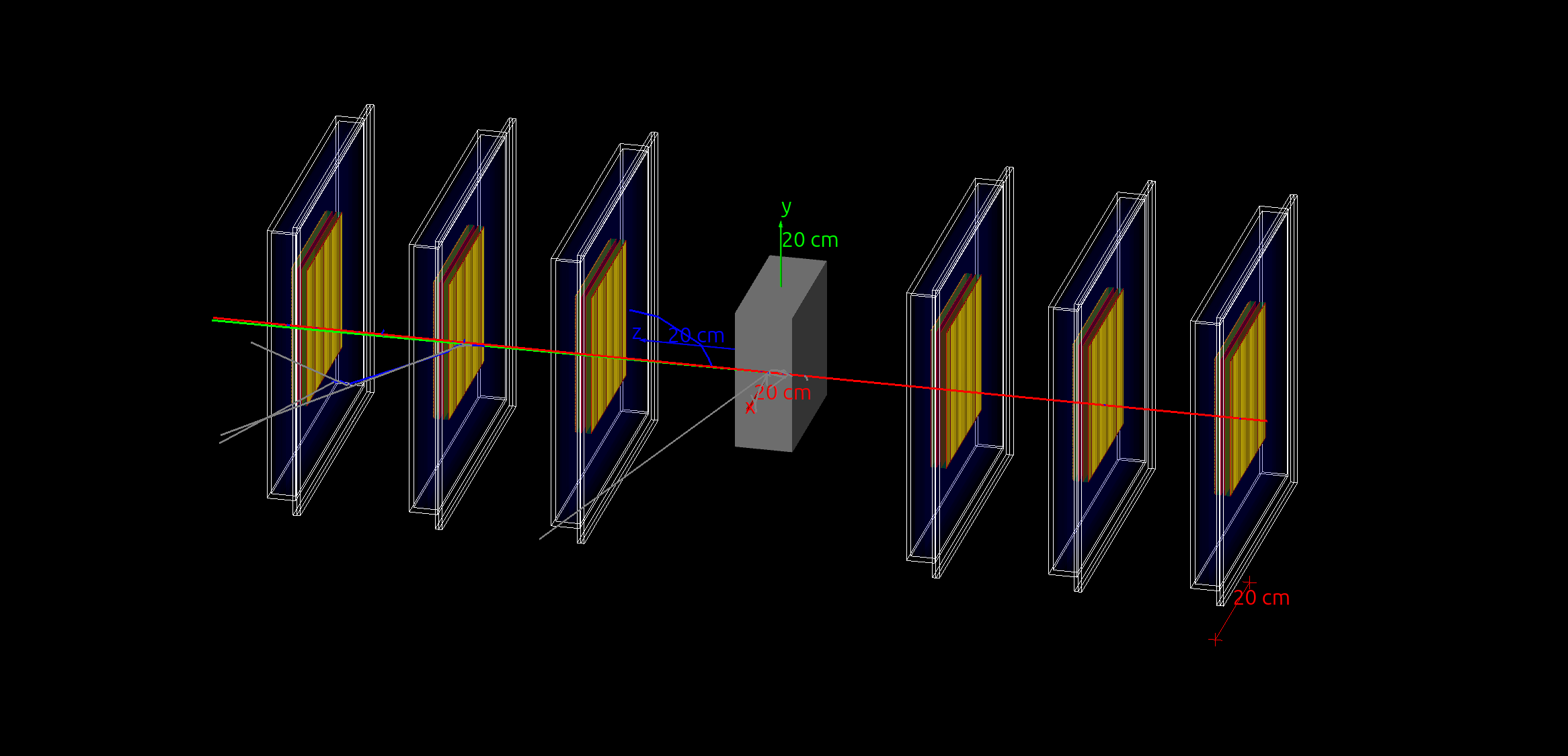}
\caption{\justifying $\mu^+e^- \to \mu^+\mu^-$ event display. The beam travels from right to left and the CLFV process occurs on the target.}
\label{fig:6RPC}
\end{figure}

In the PKMu experiment, atmospheric positive muons or positive muon beams can scatter off nuclear-bound electrons in tunable target materials. The incoming $\mu^+$ and the target $e^-$ may annihilate to produce a $Z'$ boson, which can subsequently decay into a pair of same-generation charged leptons, resulting in charged lepton flavor violation (CLFV). Typical processes include $\mu^+e^- \to \mu^+\mu^-$ and $\mu^+e^- \to e^+e^-$. By analyzing the tracks of the incoming muon and the resulting lepton pairs, we can effectively probe or constrain the parameter space of the CLFV $Z'$ model. We have conducted a preliminary feasibility study using Monte Carlo simulations~\cite{Gao:2024xvf} utilizing the \textsc{Geant4}~\cite{GEANT4:2002zbu} framework. Fig.~\ref{fig:6RPC} illustrates a typical signal event, where a muon enters from the right, interacts with an aluminum or lead target at the center, and produces two opposite-sign muons (indicated in red and green) that exit through the RPC on the left.

\begin{figure}
\centering
\subfloat[$3\ \mathrm{cm}$ Al target, with $e$-veto]{\label{fig:ul-al-ne}\includegraphics[width=.5\columnwidth]{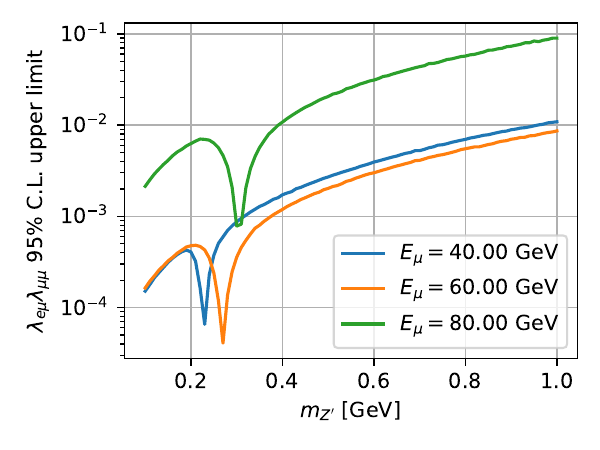}}
\subfloat[$8\ \mathrm{cm}$ Pb target, with $e$-veto]{\label{fig:ul-pb-ne}\includegraphics[width=.5\columnwidth]{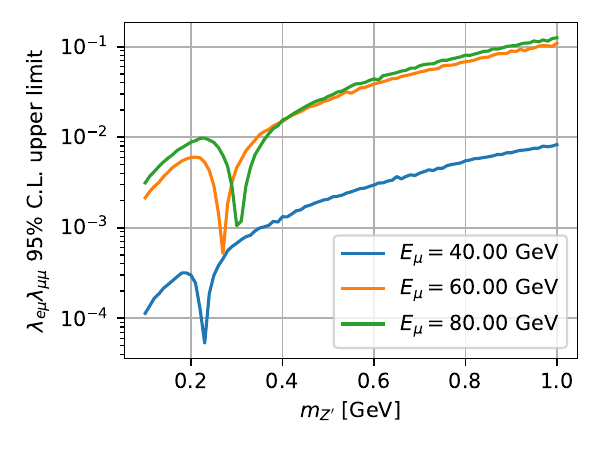}}
\caption{\justifying 95\% C.L. upper limit results of various targets. The yields are normalized to $3 \times 10^{13}$ muons on target, corresponding to a one-year run.}
\label{fig:95UL}
\end{figure}

Fig.~\ref{fig:95UL} indicates the 95\% C.L. upper limit results of various targets. The yields are normalized to $3 \times 10^{13}$ muons on target, corresponding to a one-year run. While the sensitivity of our experiment is slightly lower than that of the low-energy MEG experiment, MEG can only provide overall constraints on multiple couplings, such as the coherent interposition of $\lambda_{ee}\lambda_{e\mu}$ and $\lambda_{e\mu}\lambda_{\mu\mu}$. However, these couplings and others may exhibit significant cancellations, potentially allowing them to evade existing constraints. In contrast, our experiment can uniquely and decisively probe $\lambda_{e\mu}\lambda_{\mu\mu}$, providing a distinct advantage in exploring this parameter. On the other hand, the primary challenge in expanding this experiment lies in the electronics readout. We must efficiently distinguish between two simultaneously emitted electrons or muons and accurately measure their flight directions. The current plan includes enhancing the spatial resolution of the muon detector to 0.3–0.5 mm. Additionally, a well-designed waveform readout algorithm should be optimized to effectively decode and reconstruct the signals from closely spaced muons.

\section{Entanglement studies via lepton beam on-target experiments}

Quantum entanglement~\cite{Horodecki:2009zz} is one of the most distinctive and counterintuitive aspects of quantum mechanics. It describes a quantum system that cannot be adequately explained by classical theories and can lead to the violations of Bell inequalities~\cite{Bell:1964kc}. Experimentally, these phenomena have been successfully demonstrated using correlated photons or bound electrons. In recent years, at the high-energy frontier, the measurement and verification of quantum entanglement have become a significant focus of research. This is achieved by studying high-energy particles produced at particle colliders, including tau leptons, mesons, top quarks, and gauge bosons~\cite{Barr:2024djo}. Notably, recent announcements from the ATLAS and CMS experiments at the Large Hadron Collider (LHC) have reported the discovery of quantum entanglement between top-antitop quark pairs, marking a groundbreaking observation of quantum entanglement at the highest energy scales~\cite{ATLAS:2023fsd,CMS:2024pts,CMS:2024zkc}.

\begin{figure}
\centering
\subfloat[$\mu^-e^- \to \mu^-e^-$]{\includegraphics[width=.5\columnwidth]{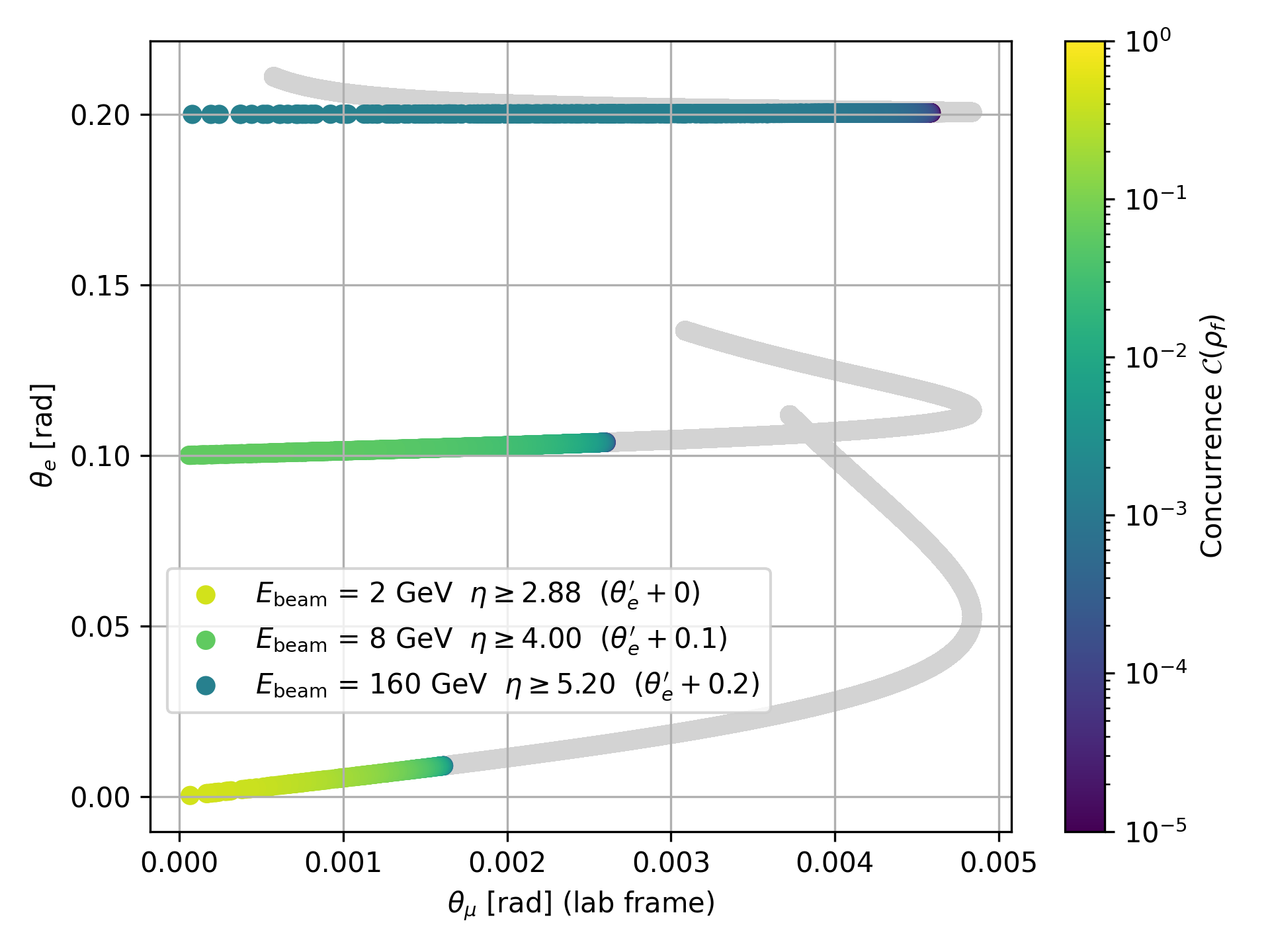}\includegraphics[width=.5\columnwidth]{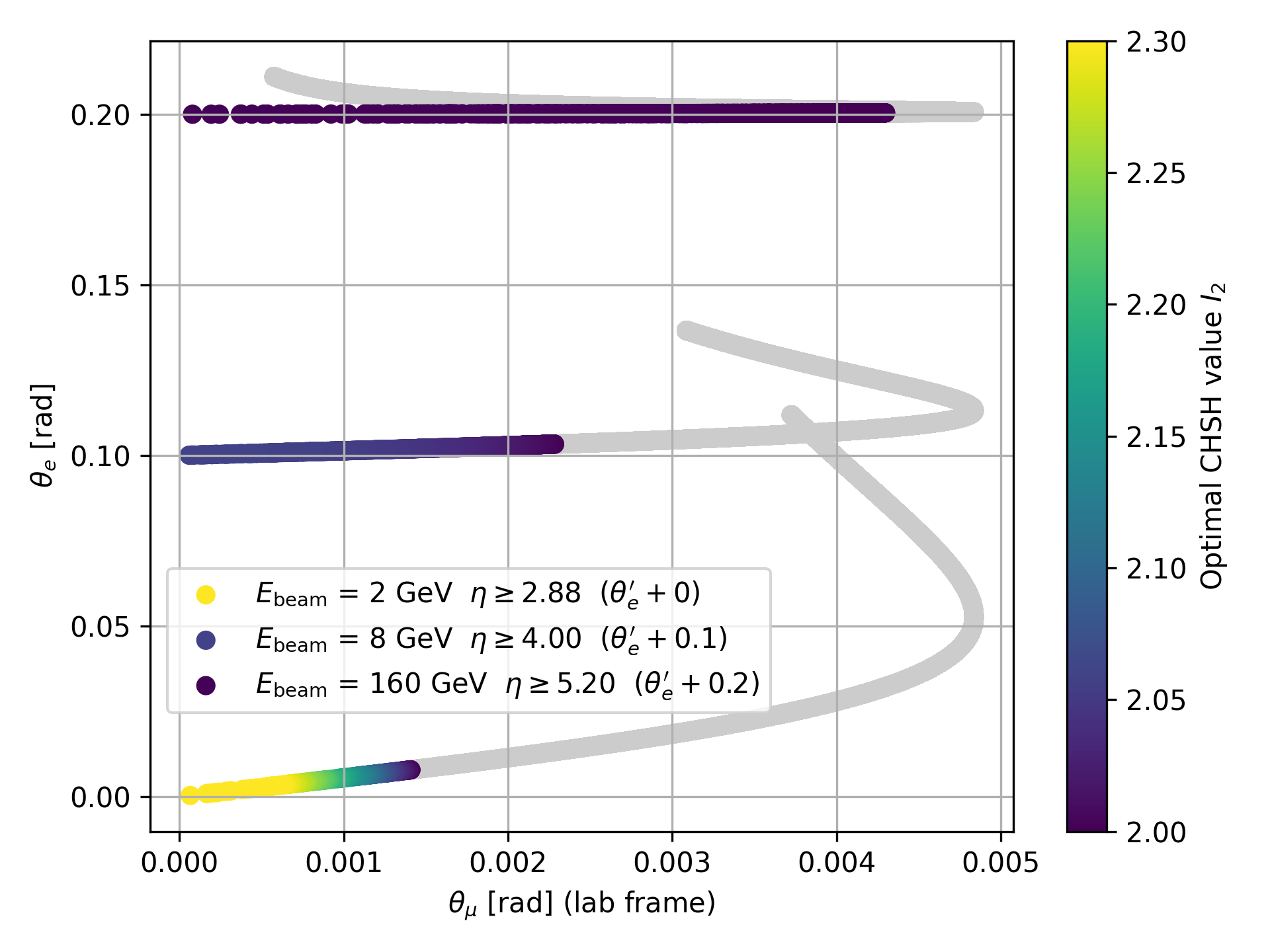}}

\subfloat[$e^+e^- \to e^+e^-$]{\includegraphics[width=.5\columnwidth]{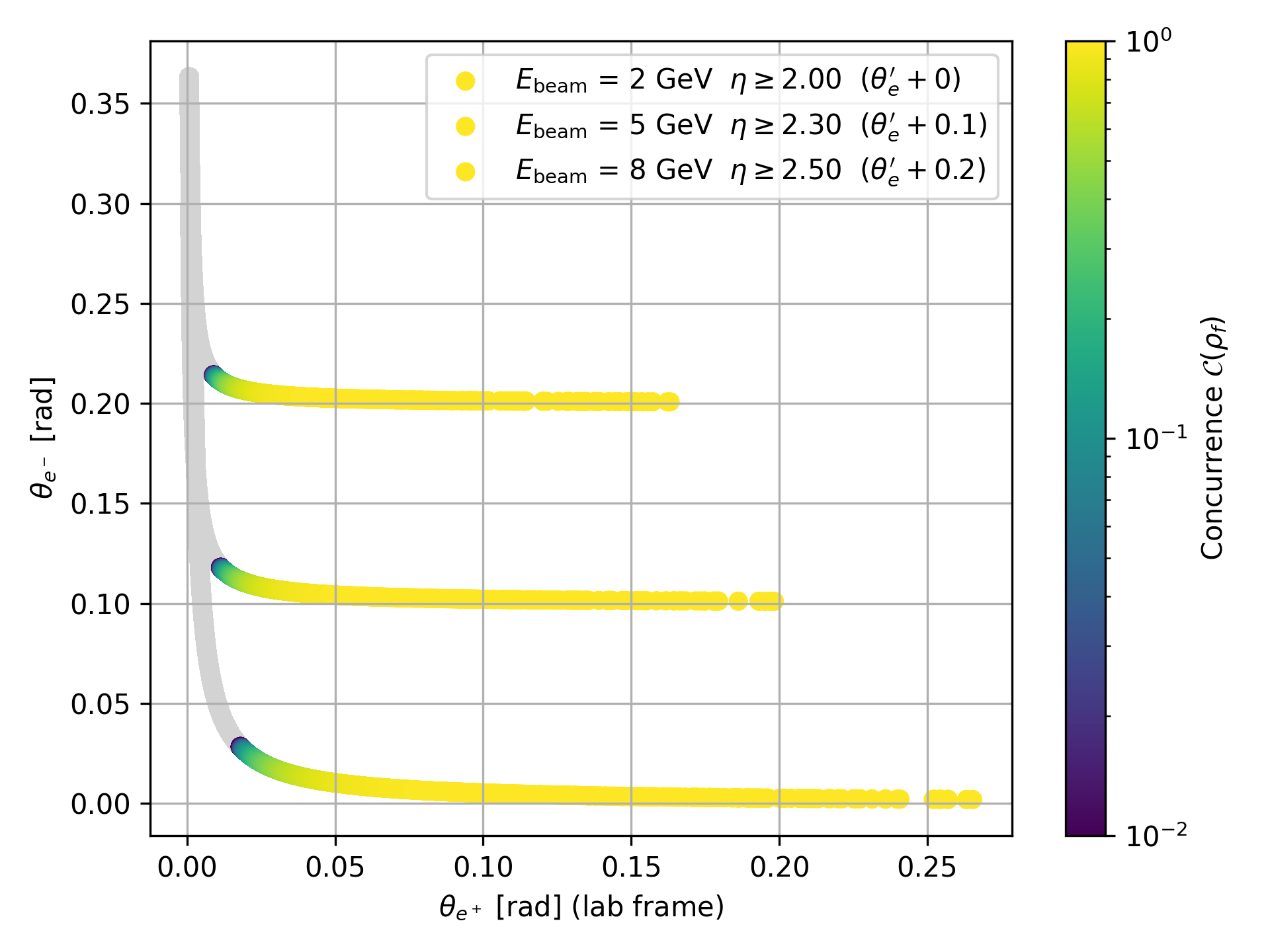}\includegraphics[width=.5\columnwidth]{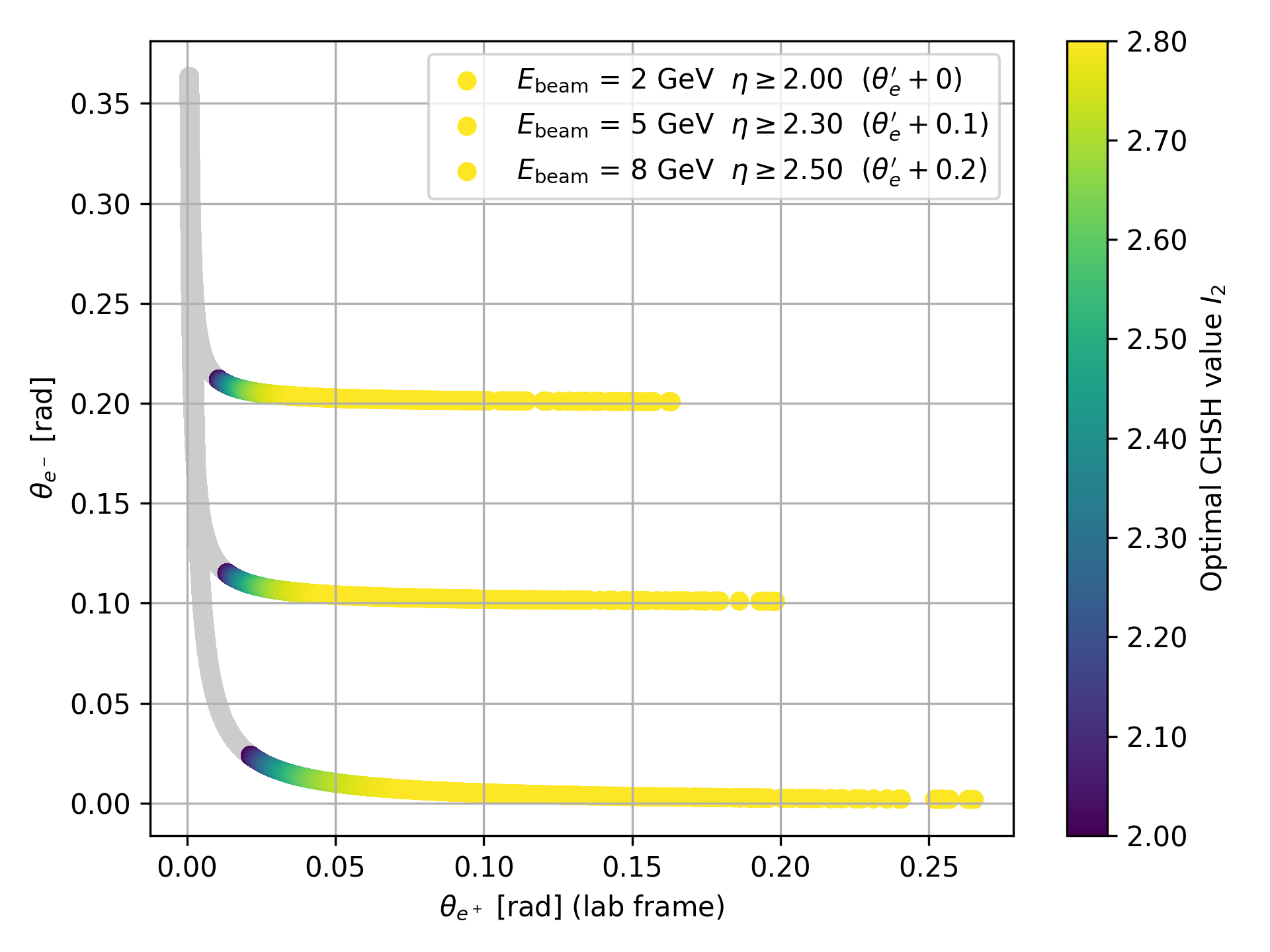}}

\caption{\justifying Final state scatter plots colored by the concurrence $\mathcal{C}(\rho_f)$ or the optimal CHSH value $I_2$. These plots correspond to various incoming muon/positron energies in the lab frame. The light gray regions indicate where $\mathcal{C}(\rho_f) = 0$ or $I_2 \leq 2$. Events are generated with the labeled minimum pseudorapidity requirements.}
\label{fig:qe}
\end{figure}

Entanglement witness quantities, such as \emph{concurrence}, can be derived from the density matrix following appropriate transformations to probe the entanglement~\cite{Gao:2024leu}. In the cases of the muon-electron and positron-electron scattering process, the density matrix can be inferred from the scattering matrix element, along with the measured momentum and angular information~\cite{Gao:2024leu}. A concurrence value greater than zero indicates the presence of quantum entanglement. In addition, the optimal CHSH value is also derivable from the density matrix to test the possible violation of the Bell inequalities for the entangled states~\cite{Gao:2024leu}. The resulting concurrence and optimal CHSH value results for the $\mu^-e^- \to \mu^-e^-$ and $e^+e^- \to e^+e^-$ processes are shown in Fig.~\ref{fig:qe}.

For $\mu^-$ on-target scattering experiments, Fig.~\ref{fig:qe} indicates that incident energies ranging from 1 to 10 GeV exhibit relatively high occurrence and $I_2 \geq 2$ values, with scattering angles predominantly distributed between 1 and 3 mrad, making it suitable for observation. And more promisingly, for $e^+$ on-target scattering, simulation results indicate a generally stronger entanglement and violations of the CHSH inequality. Quantitatively, the theoretical upper limits for both $\mathcal{C}(\rho_f)$ and $I_2$ in quantum mechanics are nearly reached in extreme cases. The event rates for the proposed experiments above are high as a result of high production cross sections. Assuming a one-day run with a 10 GeV muon beam of flux $10^5/\mathrm{s}$ on aluminum targets of $10\ \mathrm{cm}$ thick, the expected number of events with entangled final states is $2.6 \times 10^{4}$; assuming a 1 GeV positron beam with a flux of $10^{12}/\mathrm{s}$ directed at a $10\ \mathrm{cm}$ thick aluminum target, the expected entangled event rate is $1.9 \times 10^9$/s~\cite{Gao:2024leu}.

Furthermore, in the context of Bhabha scattering products, the final state can be approximated as a Bell state, in which the $z$-components of the spins of an electron-positron pair are always aligned when measured simultaneously in the lab frame. Utilizing prior knowledge of the spin state, one can measure the polarization correlations of the entangled pairs through their individual polarization-sensitive scatterings off secondary targets. Simulations of the joint distributions of the secondary scattering products, as presented in Ref.~\cite{Gao:2024leu}, demonstrate significant discrimination between different states for 100\% polarized targets, including the one between the dominating Bell state and the unpolarized state. The discriminating power remains for targets up to 20\% polarized. The measurements impose minimal statistical requirements, as statistically significant results can be obtained with just a few hundred seconds of event accumulation even for the 20\% polarized targets, when not accounting for modeling and detection uncertainties. Further details can be found in Ref.~\cite{Gao:2024leu}.

\section{Summary}

In this proposal, we introduce a novel set of methods known as probing and knocking with muons (PKMu), aimed at exploring the rich physics associated with GeV muon beams. These methods provide various avenues for investigation, including the potential for large-angle muon scattering to reveal evidence of muon-philic dark matter candidates. Additionally, muon-electron scattering can be utilized to detect new bosons linked to CLFV. Furthermore, precise measurements of GeV-scale muon-electron scattering can probe quantum correlations. These approaches may enhance our understanding of fundamental physics and uncover potential new phenomena.

\section*{Acknowledgments}

This work is supported in part by the National Natural Science Foundation of China under Grants No. 12325504.


\begin{thebibliography}{0}

\bibitem{Essig:2022dfa}
R.~Essig, G.~K.~Giovanetti, N.~Kurinsky, D.~McKinsey, K.~Ramanathan, K.~Stifter and T.~T.~Yu,
[arXiv:2203.08297 [hep-ph]].

\bibitem{Harris:2022vnx}
P.~Harris, P.~Schuster and J.~Zupan,
[arXiv:2207.08990 [hep-ph]].

\bibitem{Bai:2014osa}
Y.~Bai and J.~Berger,
JHEP \textbf{08}, 153 (2014)
doi:10.1007/JHEP08(2014)153
[arXiv:1402.6696 [hep-ph]].

\bibitem{Drees:1983pd}
J.~Drees and H.~E.~Montgomery,
Ann. Rev. Nucl. Part. Sci. \textbf{33}, 383-452 (1983)
doi:10.1146/annurev.ns.33.120183.002123

\bibitem{MEG:2016leq}
A.~M.~Baldini \textit{et al.} [MEG],
Eur. Phys. J. C \textbf{76}, no.8, 434 (2016)
doi:10.1140/epjc/s10052-016-4271-x
[arXiv:1605.05081 [hep-ex]].

\bibitem{Hesketh:2022wgw}
G.~Hesketh \textit{et al.} [Mu3e],
[arXiv:2204.00001 [hep-ex]].

\bibitem{NA64:2024klw}
Y.~M.~Andreev \textit{et al.} [NA64],
Phys. Rev. Lett. \textbf{132}, no.21, 211803 (2024)
doi:10.1103/PhysRevLett.132.211803
[arXiv:2401.01708 [hep-ex]].

\bibitem{CarloniCalame:2015obs}
C.~M.~Carloni Calame, M.~Passera, L.~Trentadue and G.~Venanzoni,
Phys. Lett. B \textbf{746}, 325-329 (2015)
doi:10.1016/j.physletb.2015.05.020
[arXiv:1504.02228 [hep-ph]].

\bibitem{McKeen:2023ztq}
D.~McKeen, D.~E.~Morrissey, M.~Pospelov, H.~Ramani and A.~Ray,
Phys. Rev. Lett. \textbf{131}, no.1, 011005 (2023)
doi:10.1103/PhysRevLett.131.011005
[arXiv:2303.03416 [hep-ph]].

\bibitem{Xu:2025spd}
Y.~Xu, X.~Zhang, Y.~Yu, P.~Yu, L.~Deng, J.~Zhai, L.~Chen, H.~Zhao, L.~Sheng and G.~Shen, \textit{et al.}
[arXiv:2502.20915 [physics.acc-ph]].

\bibitem{Ruzi:2023mxp}
A.~Ruzi, C.~Zhou, X.~Sun, D.~Wang, S.~Wang, Y.~Ban, Y.~Mao, Q.~Li and Q.~Li,
Int. J. Mod. Phys. A \textbf{38} (2023) no.29n30, 2350154
doi:10.1142/S0217751X23501543
[arXiv:2303.18117 [hep-ph]].

\bibitem{Yu:2024spj}
X.~Yu, Z.~Wang, C.~e.~Liu, Y.~Feng, J.~Li, X.~Geng, Y.~Zhang, L.~Gao, R.~Jiang and Y.~Wu, \textit{et al.}
Phys. Rev. D \textbf{110}, no.1, 016017 (2024)
doi:10.1103/PhysRevD.110.016017
[arXiv:2402.13483 [hep-ex]].

\bibitem{Gao:2024xvf}
L.~Gao, Z.~Wang, C.~e.~Liu, J.~Li, A.~Ruzi, Q.~Li, C.~Zhou and Q.~Li,
[arXiv:2410.20323 [hep-ex]].

\bibitem{Gao:2024leu}
L.~Gao, A.~Ruzi, Q.~Li, C.~Zhou, L.~Chen, X.~Zhang, Z.~Sun and Q.~Li,
[arXiv:2411.12518 [hep-ph]].

\bibitem{Gao:2025kdi}
L.~Gao, A.~Ruzi, Q.~Li, C.~Zhou and Q.~Li,
[arXiv:2502.07597 [hep-ph]].

\bibitem{Abbas:2022fze}
M.~Abbas, M.~Abbrescia, H.~Abdalla, A.~Abdelalim, S.~AbuZeid, A.~Agapitos, A.~Ahmad, A.~Ahmed, W.~Ahmed and C.~Aim\`e, \textit{et al.}
Nucl. Instrum. Meth. A \textbf{1034}, 166716 (2022)
doi:10.1016/j.nima.2022.166716
[arXiv:2203.12037 [physics.ins-det]].

\bibitem{Pellecchia:2022lsd}
A.~Pellecchia \textit{et al.} [CMS Muon],
Nucl. Instrum. Meth. A \textbf{1046}, 167618 (2023)
doi:10.1016/j.nima.2022.167618
[arXiv:2207.09906 [physics.ins-det]].

\bibitem{GEANT4:2002zbu}
S.~Agostinelli \textit{et al.} (GEANT4 Collaboration),
Nucl. Instrum. Meth. A \textbf{506}, 250-303 (2003)

\bibitem{gemd}
https://ep-news.web.cern.ch/content/gems-cms

\bibitem{Riegler:2004gc}
W.~Riegler and C.~Lippmann,
Nucl. Instrum. Meth. A \textbf{518}, 86-90 (2004)
doi:10.1016/j.nima.2003.10.031

\bibitem{Lewin:1995rx}
J.~D.~Lewin and P.~F.~Smith,
Astropart. Phys. \textbf{6} (1996), 87-112
doi:10.1016/S0927-6505(96)00047-3

\bibitem{Aalbers:2018mfc}
J.~Aalbers,
en. 9789462339842, 2018. ISBN: 978-94-6233-984-2

\bibitem{CRY}
C.~Hagmann, D.~Lange, D.~Wright
IEEE Nuclear Science Symposium Conference Record.IEEE, doi:10.1109/NSSMIC.2007.4437209.

\bibitem{Battaglieri:2017aum}
M.~Battaglieri, A.~Belloni, A.~Chou, P.~Cushman, B.~Echenard, R.~Essig, J.~Estrada, J.~L.~Feng, B.~Flaugher and P.~J.~Fox, \textit{et al.}
[arXiv:1707.04591 [hep-ph]].

\bibitem{Foot:1990mn}
R.~Foot,
Mod. Phys. Lett. A \textbf{6}, 527-530 (1991)
doi:10.1142/S0217732391000543

\bibitem{He:1990pn}
X.~G.~He, G.~C.~Joshi, H.~Lew and R.~R.~Volkas,
Phys. Rev. D \textbf{43}, 22-24 (1991)
doi:10.1103/PhysRevD.43.R22

\bibitem{He:1991qd}
X.~G.~He, G.~C.~Joshi, H.~Lew and R.~R.~Volkas,
Phys. Rev. D \textbf{44}, 2118-2132 (1991)
doi:10.1103/PhysRevD.44.2118

\bibitem{Foot:1994vd}
R.~Foot, X.~G.~He, H.~Lew and R.~R.~Volkas,
Phys. Rev. D \textbf{50}, 4571-4580 (1994)
doi:10.1103/PhysRevD.50.4571
[arXiv:hep-ph/9401250 [hep-ph]].

\bibitem{wang2025}
Zijian Wang, \textit{et al.}, to appear soon.

\bibitem{Ding:2024zaj}
R.~Ding, J.~Li, M.~Lu, Z.~You, Z.~Wang and Q.~Li,
JHEP \textbf{01}, 165 (2025)
doi:10.1007/JHEP01(2025)165
[arXiv:2405.09417 [hep-ex]].

\bibitem{Langacker:2000ju}
P.~Langacker and M.~Plumacher,
Phys. Rev. D \textbf{62}, 013006 (2000)
doi:10.1103/PhysRevD.62.013006
[arXiv:hep-ph/0001204 [hep-ph]].

\bibitem{Alwall:2014hca}
J.~Alwall, R.~Frederix, S.~Frixione, V.~Hirschi, F.~Maltoni, O.~Mattelaer, H.~S.~Shao, T.~Stelzer, P.~Torrielli and M.~Zaro,
JHEP \textbf{07}, 079 (2014)
doi:10.1007/JHEP07(2014)079
[arXiv:1405.0301 [hep-ph]].

\bibitem{Horodecki:2009zz}
R.~Horodecki, P.~Horodecki, M.~Horodecki and K.~Horodecki,
Rev. Mod. Phys. \textbf{81}, 865-942 (2009)
doi:10.1103/RevModPhys.81.865
[arXiv:quant-ph/0702225 [quant-ph]].

\bibitem{Bell:1964kc}
J.~S.~Bell,
Physics Physique Fizika \textbf{1}, 195-200 (1964)
doi:10.1103/PhysicsPhysiqueFizika.1.195.

\bibitem{Barr:2024djo}
A.~J.~Barr, M.~Fabbrichesi, R.~Floreanini, E.~Gabrielli and L.~Marzola,
Prog. Part. Nucl. Phys. \textbf{139}, 104134 (2024)
doi:10.1016/j.ppnp.2024.104134
[arXiv:2402.07972 [hep-ph]].

\bibitem{ATLAS:2023fsd}
G.~Aad \textit{et al.} [ATLAS],
Nature \textbf{633}, no.8030, 542-547 (2024)
doi:10.1038/s41586-024-07824-z
[arXiv:2311.07288 [hep-ex]].

\bibitem{CMS:2024pts}
A.~Hayrapetyan \textit{et al.} [CMS],
Rept. Prog. Phys. \textbf{87}, no.11, 117801 (2024)
doi:10.1088/1361-6633/ad7e4d
[arXiv:2406.03976 [hep-ex]].

\bibitem{CMS:2024zkc}
A.~Hayrapetyan \textit{et al.} [CMS],
Phys. Rev. D \textbf{110}, no.11, 112016 (2024)
doi:10.1103/PhysRevD.110.112016
[arXiv:2409.11067 [hep-ex]].

\end{thebibliography}
\end{document}